\documentclass[conference]{IEEEtran}
\IEEEoverridecommandlockouts

\usepackage{comment}
\usepackage{graphicx}
\usepackage{amsmath}
\usepackage{url}
\usepackage{threeparttable} 
\usepackage{booktabs}
\usepackage{multirow}
\usepackage{listings}
\usepackage{xcolor}
\usepackage{float}
\usepackage{url}
\usepackage{hyperref}
\usepackage{xurl}
\usepackage{balance}
\usepackage{array}
\usepackage{makecell}  

\usepackage{pgfplots}
\pgfplotsset{compat=1.18} 

\usepackage{tikz}
\usetikzlibrary{arrows.meta,positioning,shapes.geometric,shapes.misc,fit}

\tikzset{
  >=Latex,
  flow/.style={line width=0.6pt},
  term/.style={ellipse, draw, flow, align=center, minimum width=17mm, minimum height=8mm, fill=white},
  proc/.style={rectangle, rounded corners=2pt, draw, flow, align=center, minimum width=28mm, minimum height=8mm, fill=white},
  decision/.style={diamond, aspect=2, draw, flow, align=center, inner ysep=2pt, minimum width=22mm, fill=white},
  note/.style={rectangle, draw, flow, align=center, font=\scriptsize,
               rounded corners=2pt, inner sep=2pt, fill=white},
  groupbox/.style={draw, rounded corners=2pt, flow, inner sep=3pt},
  lab/.style={font=\scriptsize, inner sep=1pt, fill=white}
}

\begin{document}

\title{Safe2Hail: A Forensic-Driven Post-Trip Tracking Framework for Ride-Hailing Safety in Africa}

\author{
\IEEEauthorblockN{Alvina  Minja, Robert Maina, Mahmud Oloyede, Jema Ndibwile\textsuperscript{*}}
\IEEEauthorblockA{College of Engineering, Carnegie Mellon University Africa, Kigali, Rwanda\\
Email: \{aminja, rmaina, moloyede, jndibwil\}@andrew.cmu.edu}
\IEEEauthorblockA{*Corresponding author}
}

\maketitle

\begin{abstract}
Ride-hailing mobile apps have become an essential feature in the mobility ecosystem in Africa, offering much safer and much more affordable rides. Although user bases have increased and the number of daily trips has proliferated, reports of imminent safety threats, particularly after the cancellation of the ride or the ride is prematurely terminated, remain unresolved challenges. Current safety measures offer features such as SOS alerts, safety notifications, and live location sharing for the duration of the trip, but they are not in place when the trip is over. Safe2Hail presents a framework that is forensically driven to ensure continuous safety and certainty beyond the trip. The Safe2Hail framework combines forensic tracking with a temporary post-trip synchronization mechanism that can securely log all proximal data between a passenger and a driver after an event. The Safe2Hail framework was beta-tested and demonstrates the effectiveness of the system. Although the research team did not pilot on actual deployment, the Safe2Hail design format was in part inspired by actual crime events reported in Nairobi and Dar-es-Salaam. The findings of the study referenced Safe2Hail's feasibility, light weight nature of resources, as well as the scalability for the framework.
\end{abstract}


\section{Introduction} 
Ride-hailing services have become essential in contemporary African cities by altering transportation habits, work patterns, and the way we share transportation. However, such rapid digitization has led to complicated safety and accountability problems that exceed the traditional scope of monitoring the trip itself. This section describes the contextual background behind the study to situate these problems and the proposed intervention, identifies the central problem being addressed by the intervention, and summarizes the contributions of these research findings.

\subsection{Context and Motivation}
The ride-hailing industry has, since its inception, offered convenience in the transport industry, growing rapidly daily, especially in urban cities. The Middle East \& Africa Ride Hailing market size is expected to reach USD 9,992 Million by 2033 \cite{Spherical2024}. This growth is attributed not only to its convenience but the job offers that come along with it, and the need for safe travel. This sector was valued at \$345 million in 2023, employing over 93,000 workers across 23 companies, with an adoption rate of 46\% \cite{omolo_digital_2024}. Some of the key players in the market include Uber, Bolt (Taxify), and inDrive, which also operate in our countries of interest, Tanzania and Kenya. But while the need for this service grows, so do its safety risks. 

The sector faces persistent challenges, particularly regarding user safety and security. Security concerns have become a critical factor in the adoption and reliability of ride-hailing services.  A study conducted by \textit{Deutsche Gesellschaft für Internationale Zusammenarbeit} (GIZ)
revealed that users prioritized security and safety (78\%) and comfort (70\%) as decisive factors in their transportation choices \cite{Weru2020}.  
Recent studies on ride-hailing safety in East Africa reveal recurring incidents of robbery, abduction, and sexual harassment, particularly among urban riders in Nairobi and Dar es Salaam. These safety concerns have led some companies to withdraw services from high-risk
areas, further highlighting the urgent need for improved security measures \cite{Weru2020}\cite{kamais2019emerging}.

Although several initiatives and safety features have been implemented in ride-hailing services to reduce crime, the specific risks associated with trip cancellations, premature ride endings, and ill-intended destination changes remain underexplored. These trip-level anomalies present significant safety concerns that warrant closer investigation that this research aims to uncover, these risks leave passengers stranded in unsafe locations, deprive drivers of income opportunities, and impede incident(forensic) investigations. 

Recognizing the significant growth and inherent safety challenges within the ride-hailing sector across East Africa, this research specifically focuses on Tanzania and Kenya. These nations serve as critical markets for major ride-hailing platforms such as Uber, Bolt, and inDrive, which have established substantial operations in both countries \cite{mwangi_ride_2025}\cite{noauthor_revolutionizing_nodate}. A research by Sagaci indicates that Kenya alongside South Africa and Nigeria, exhibits some of the highest ride-hailing app usage penetration across the continent, with Bolt being the most used mobility app in Tanzania \cite{pioch_revolutionising_2023}. Furthermore, both countries, particularly urban centers such as Nairobi and Dar es Salaam, have seen a growing number of reports of safety incidents, including trip cancellations, premature ride endings, and unauthorized destination changes, as highlighted by various news outlets and studies \cite{noauthor_revolutionizing_nodate}\cite{attorneys_breakthrough_2022}. This study aims to provide a more in-depth and actionable understanding of trip-level anomalies and their impact on user safety and forensic investigations within a localized but indicative regional setting.

\subsection{Problem Statement}
Passengers, particularly women and informal workers, continue to face serious criminal incidents including robbery, harassment, kidnapping, and fatal assault despite the widespread use of services like Uber and Bolt in East African cities.  Trip sharing, GPS monitoring, and SOS warnings are some of the safety features, but once a trip is canceled or ends prematurely, passengers no longer have that protection. This leaves a critical forensic gap, as potential crimes occur beyond the reach of platform monitoring.

To address this, the present study develops a technical intervention that enforces trip integrity, maintains forensic traceability beyond the trip endpoint, and supports standardized post-incident investigation across ride-hailing services.

\subsection{Contribution}
Building on prior studies of ride-hailing safety, this work introduces Safe2Hail, a forensic-aware framework that enhances post-trip accountability. The key contributions are:

\begin{enumerate}
\item \textbf{Post-Track Mechanism:} A lightweight module that temporarily extends monitoring after trip anomalies to preserve verifiable digital trails while maintaining privacy.

\item \textbf{Contextual Risk Mapping}: Using a systematic risk matrix, this approach systematically classifies and ranks trip anomalies (such as GPS spoofing and premature trip terminations) that occur in Tanzania and Kenya.

\item \textbf{Dynamic Risk Multipliers}: Using contextual factors (such as crime rates, driving records, and the time of day) to enhance the score of detected irregularities and elevate high-risk instances is known as dynamic risk multipliers.
\end{enumerate}

\section{Literature Review} 

\subsection{Threats in Ride Hailing}
The ride-hailing industry in East Africa, particularly in Kenya and Tanzania, has experienced significant growth and transformation. However, it also faces several threats that could jeopardize its sustainability and impact. In this section of the literature review, we focus on the findings concerning the current threats facing ride-hailing services in this region, focusing on operational challenges, safety concerns, regulatory dilemmas, and socio-economic impacts.

Among the numerous threats and risks in ride hailing services, safety concerns for both drivers and passengers have emerged as a critical issue that requires urgent solutions. Although most studies focus on passenger security, it is high time we also investigate the security concerns faced by drivers. The risky nature of gig work that comes with ride hailing services exposes drivers to threats such as violence, harassment, and robbery. A study carried out by Anwar et al. reveals that drivers in the ride-hailing sector in Kenya report various risks and often feel unprotected due to lack of robust security protocols \cite {Anwar.2022}. This concern is supported by Lefcoe et al., in a finding that ride-hailing drivers may carry weapons for self-defense due to inadequate protective measures from the companies they work for \cite {Lefcoe.2023}. In addition, the issue of gender safety is pronounced with women reporting a heightened sense of vulnerability when using ride-hailing services \cite {hu2024safety}.

On the other hand, physical security risks are among the most concerning issues in ride-hailing services for most passengers. In a study conducted focusing on Uber Taxi in Kenya, it revealed that the risks perceived as most likely include robbery (41.84\%), abductions (40.82\%), carjacking (40.82\%), sexual harassment (38.14\%), murders (35.71\%), and burglaries (34.69\%) [8]. Interestingly, 28.57\% of the respondents thought that hacking into sensitive customer and company data was less likely \cite {kamais2019emerging}. The authors suggest this low perception of hacking risk might be due to a lack of public information about such incidents, which companies might keep internal for damage control, or because such incidents are genuinely rare.

\subsection{Security Measures in Place}
Governments, NGOs, and ride-hailing service providers such as Uber and Bolt have made several efforts to suppress the security concerns arising in this sector. Given the novel nature of ride-hailing security concerns, governments have had to draft new laws and regulatory frameworks to address the concerns. In one study conducted in Tanzania, the authors emphasize that favorable policies and regulatory compliance are essential for ensuring secure ride-hailing operations in Tanzania \cite {masele2024determinants}. These regulations often mandate specific safety protocols that ride-hailing companies must implement, such as background checks for drivers and passenger safety education programs. Such measures enhance accountability and can mitigate risks associated with criminal activities during rides.

Driver training and user education have been cited as top strategies for minimizing security and safety concerns in ride-hailing services. Driver training and emergency protocols have been deemed pivotal. These programs focus on equipping drivers with skills to handle emergencies and conflict situations. For instance, practices encompassing communication with emergency services and conflict de-escalation techniques \cite {afifudin2024driver}. Effective training promotes confidence among drivers and instills a sense of security for riders.

Additionally, studies have shown that having a well-structured emergency response system in place can significantly elevate the perceived safety of both drivers and passengers \cite {afifudin2024driver}. On the other hand, passenger training has also been studied as an important aspect in ride-hailing security. Passengers are often encouraged to engage in practices that enhance their safety, such as verifying vehicle details and driver identity before commencing rides. Hu and Yang \cite {hu2024safety} propose strategies to enhance user awareness of safety protocols, particularly focusing on female passengers who may have heightened security concerns. This study concludes that initiatives that educate users about recognizing safe and dangerous situations contribute to a more secure environment for all participants in the ride-hailing system. 

Ride hailing service providers have the obligation to ensure drivers and passengers are safe when using their services. One of the approaches that the service providers have used is the implementation of SOS buttons that alert emergency services or predefined contacts during emergencies. Hu and Yang emphasize that such features, especially designed with female passengers in mind, enable swift emergency responses, which are crucial given the vulnerabilities reported by women in ride-hailing scenarios \cite{hu2024safety}. Additionally, the integration of real-time GPS tracking allows both passengers and drivers to monitor routes in real-time, substantially reducing the chances of kidnapping or falling victim to improperly vetted drivers. In response to gender-based safety concerns, some service providers have implemented innovative solutions. For instance, Bolt introduced a women-only service in Nairobi \cite{washingtonpost2021kenya3}. However, this initiative faced significant challenges due to higher pricing and limited driver availability, ultimately failing to address the core safety issues. This outcome connects with broader findings about the need for comprehensive safety solutions rather than gender-specific approaches. 

\subsection{Technological Solutions for Ride-Hailing Safety}
Rising security and safety concerns with ride-hailing services have led to attempts to design technology-based countermeasures for passenger safety and care, as well as for forensic accountability at present. Various studies emphasize the automated design, real-time monitoring systems (to track trajectories), and data privacy methods for countering ride-hailing concerns.

An example of a framework developed is the Online Security Passenger Protection (onSecP) system that provides smooth integration of human geographic data with trajectory analysis to assess ride-hailing safety risks. The framework improves situational awareness through real-time route (trajectory) monitoring and acts as a deterrent for would-be assailants. Though there are monitoring systems, trajectory analysis techniques, and techniques for assessing risk that are well known and published, the development of a monitoring system that incorporates human geographic information and trajectory monitoring, as demonstrated in onSecP remains an advancement to mobility security. The onSecP system merges a Longest Common Subsequence (LCSS)-K-means-geoinformation model for trajectory clustering with an Analytic Hierarchy Process (AHP)-Entropy-Cluster weighting process for risk assessments. This new hybrid design addresses significant shortcomings in the existing literature and contributes a comprehensive and data-driven safety standard for passenger care and safety as a part of mobility services \cite{fu2025monitoring}.

Research on ride-hailing has not fully addressed data privacy, but a number of papers suggest ways to enhance identity and location protection. One study, for instance, proposes a decentralized identity management framework that separates user identity from trip information, which is expected to enhance privacy and trust in the system \cite{sun2024decoupling}. This uses Decentralized Identifiers (DIDs) that are authenticated through government-issued Verifiable Credentials (VCs), published on the InterPlanetary File System (IPFS) and the associated Content Identifiers (CIDs) anchored on a blockchain. The paper applies fuzzy Point-of-Interest (POI) `matching' and Ciphertext-Policy Attribute-Based Encryption (CP-ABE) to hide the user's precise location and apply attribute-based access control. It applies Advanced Encryption Standard (AES) over CP-ABE in order to offer privacy as well as strength in cryptography. In total, all these abstraction levels protect the user's identity as well as spatial privacy even while completing ride-hailing operations successfully.

Solutions like SOS buttons, GPS tracking, and anomaly detection \cite{hu2024safety}\cite{afifudin2024driver}, as well as sophisticated frameworks like onSecP \cite{fu2025monitoring} and decentralized identity schemes \cite{sun2024decoupling} enhance passenger safety and privacy during rides. None, however, covers the crucial forensic traceability gap that appears after a trip is prematurely canceled, terminated, or changed. This oversight hinders investigative procedures and exposes drivers and passengers to risk in post-trip situations. No mechanism has been proposed in existing work on ride-hailing security that extends monitoring past the official end of a trip. By implementing a forensic-aware post-track mechanism that maintains a verifiable trail of interactions, this study closes the gap and advances ride-hailing safety research.

\section{Methodology}

This section outlines the methodological approach adopted in this study. It describes the overall design, system modeling, and analytical procedures used to develop and evaluate the proposed post-track framework. The methodology establishes the procedural steps through which the research objectives are achieved and provides the foundation for the subsequent system design.

\subsection {System Design}
\begin{figure}[!t]
    \centering
\includegraphics[width=\linewidth]{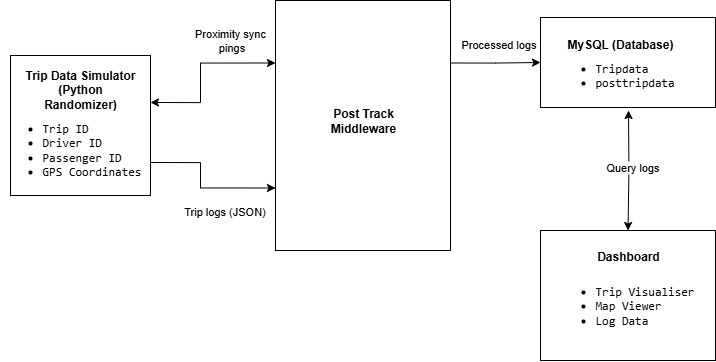}
    \caption{Block Diagram of the Safe2Hail System}
    \label{fig:fig1}
\end{figure}

Our proposed system is a modular forensics-aware platform designed to simulate a ride hailing trip with trip anomaly trigger and initiate a post-track process. It consists of three core components: (1) the trip ride data simulator, (2) the post-track engine, and (3) the dashboard as shown in Fig.~\ref{fig:fig1}. The post-track engine receives trip telemetry and is initiated upon detection of a trip anomaly. The dashboard visualizes the different ride hailing trips and shows the trace of post-track logs. The post-trip proximity sync logic improves the reliability of the evidence in the event of a premature trip ending.

\subsubsection{Trip and Trip Anomalies}
Within the boundaries of this research project, a trip is defined as a passenger initiated a trip request through a ride-hail application that is made up of a driver and consists of pickup, transit, and drop-off, as tracked through GPS, timestamp, and user incidental. Trip anomalies refer to atypical or suspicious events or patterns seen within a trip that do not follow the boundaries of typical trip operations and imply a safety concern, a potential fraud, or service disruption.
In this study, we initially identified 13 trip anomalies including:  premature trip endings, where the driver or passenger cancels after the passenger has boarded or after the driver has started the trip; significant route deviations; trips ending in a different zone from the intended destination; long and unplanned mid-trip stops; frequent driver cancellations; repeated passenger cancellations; SOS triggers during a trip caused by abnormal driver behavior (e.g., sudden braking or acceleration correlated with distress signals); multiple accounts used for the same driver or passenger; stops at high risk zones; changes to a closer drop-off point compared to the original destination; GPS anomalies (e.g., driver disables GPS mid-trip or “ghost trips” where a trip is logged without GPS movement); frequent changes of drop-off locations; and speed violations.

\subsubsection{Risk Score Calculation Methodology}
The risk score is a composite value designed to provide a real-time assessment of the severity of a security threat. The scores and multipliers in the risk matrix were established through a rigorous process that integrated empirical data from ride-hailing safety reports, crime statistics specific to Nairobi (Kenya) and Dar es Salaam (Tanzania), academic studies on risk factors in transportation, and consultations with local safety experts and ride-hailing platform representatives. This approach ensured that the assignments were evidence-based, reflecting both quantitative incident rates and qualitative insights into how environmental, behavioral, and temporal factors amplify safety risks in the ride-hailing context. We drew from publicly available safety reports from platforms like Uber and Bolt, government travel advisories, and peer-reviewed research to quantify relative risks on a normalized scale. We calculated the risk scores as shown in Equation 1:


\begin{equation}
R = C \times R_h \times T_d
\label{eq:risk}
\end{equation}

where $R$ is the overall risk score, $C$ is the crime-level score, $R_h$ is the rating-history multiplier, and $T_d$ is the time-of-day multiplier.

The components of this formula are derived from a pre-defined risk matrix, which assigns values to each factor. This methodology ensures the risk matrix is robust, adaptable, and grounded in verifiable data, allowing for accurate anomaly severity assessments in the Safe2Hail system. Table~\ref{tab:risk_factors} depicts the scoring matrix that we used to determine the base values and multipliers for each factor.

\begin{table}[!t]
\caption{Risk Factor Assessment and Scoring}
\label{tab:risk_factors}
\centering
\footnotesize 
\setlength{\tabcolsep}{2.6pt} 
\renewcommand{\arraystretch}{1.05} 
\begin{tabular}{p{1.1cm}p{1.4cm}p{0.6cm}p{3.0cm}}
\toprule
\textbf{Factor} & \textbf{Sub-Factor} & \textbf{Val} & \textbf{Rationale} \\
\midrule
\multirow{3}{*}{Crime Level} 
& High Area & 5 & Higher probability of malicious intent.\\
& Medium Area & 3 & Increased risk requires caution.\\
& Low Area & 1 & Minimal environmental risk.\\
\midrule
\multirow{3}{*}{Rating History}
& Poor & 2.5× & Negative or potentially unsafe behaviour.\\
& Neutral & 1.0× & Standard profile, no major issues.\\
& Good & 0.5× & Positive record reduces risk.\\
\midrule
\multirow{3}{*}{Time of Day}
& Late Night & 2.0× & High crime rates, low visibility.\\
& Evening & 1.5× & Common time for incidents.\\
& Day & 1.0× & Lowest risk period.\\
\bottomrule
\end{tabular}
\end{table}

\subsection{Development of the Anomaly Risk Matrix} The anomaly risk matrix, which evaluates trip anomalies based on likelihood (frequency of occurrence) and impact (severity of consequences), was developed to prioritize safety risks in ride-hailing services, particularly in high-penetration areas like Nairobi (Kenya) and Dar es Salaam (Tanzania). The risk score for each anomaly is calculated as the product of its likelihood and impact scores, both rated on a 1-5 scale, where 1 indicates low probability/minimal harm and 5 denotes high frequency/severe danger. This multiplicative approach emphasizes anomalies that are both common and highly damaging, enabling targeted interventions in the Safe2Hail system. The top anomalies (e.g., SOS triggers with a score of 25) were selected for focus based on these scores, informing the system's post-track engine.
Scores were derived through a multi-step, evidence-based process integrating quantitative data from ride-hailing safety reports, crime statistics, academic studies, and expert consultations. We prioritized regional data from East Africa to ensure relevance, drawing from sources like Uber's global and U.S. safety reports (adapted for analogous patterns in Africa), Bolt's safety initiatives in Kenya, and local incident analyses. The process involved:
\begin{enumerate}
    \item Data Collection and Review: We aggregated incident data from platform reports, government advisories, and peer-reviewed studies on ride-hailing risks in Nairobi and Dar es Salaam. For example, Uber's safety reports provided benchmarks on assault frequencies and anomaly types, while Bolt's regional statistics highlighted offline trips and driver non-compliance as proxies for anomalies like premature endings and GPS issues. Local studies emphasized emerging risks such as route deviations leading to assault.
    \item Likelihood Assessment: Likelihood scores (1-5) were assigned based on reported frequencies from historical data. High-likelihood anomalies (e.g., 5) were those occurring in a significant proportion of trips, validated by platform metrics showing, for instance, thousands of serious incidents annually. We normalized frequencies against total rides (e.g., Uber's 1.8 billion trips with 0.0002\% dangerous incidents) and adjusted for East African contexts where robbery and harassment are more prevalent \cite{mageto2025perceived}.
    \item Impact Assessment: Impact scores (1-5) reflected potential harm, including physical injury, financial loss, or psychological trauma, drawn from case studies and severity classifications in safety reports. Severe impacts (e.g., 5) were linked to life-threatening outcomes like assaults, while lower scores applied to disruptions with minimal harm.
    \item Validation and Scoring: Scores were cross-validated with experts from ride-hailing operators such as Bolt's safety team in Tanzania and refined using regression analyses from studies correlating anomalies with incidents. 
\end{enumerate}


\begin{table}[!t]
\caption{Trip-Level Risk Event Scoring}
\label{tab:table2}
\centering
\scriptsize
\setlength{\tabcolsep}{2.6pt}
\renewcommand{\arraystretch}{1.08}

\newcolumntype{L}[1]{>{\raggedright\arraybackslash}p{#1}}
\newcolumntype{C}[1]{>{\centering\arraybackslash}p{#1}}

\begin{threeparttable}
\begin{tabular}{L{3.55cm} C{1.05cm} C{1.05cm} C{1.25cm}}
\toprule
\textbf{Event} & \textbf{Lik.} & \textbf{Imp.} & \textbf{Score} \\
\midrule
SOS Triggered & 5 & 5 & 25 \\
Premature Trip Ending & 5 & 4 & 20 \\
GPS Anomalies & 4 & 4 & 16 \\
Significant route deviations & 5 & 3 & 15 \\
Frequent stops in high-risk zones & 3 & 5 & 15 \\
Multiple accounts for the same driver or passenger & 4 & 3 & 12 \\
Trip ending in a different zone from the intended destination & 3 & 4 & 12 \\
Long and unplanned mid-trip stops & 5 & 2 & 10 \\
Frequent changes of drop-off location & 3 & 3 & 9 \\
Speed violations & 4 & 2 & 8 \\
Frequent trip cancellations by the driver & 2 & 2 & 4 \\
Changes to a closer drop-off point compared to original destination & 3 & 1 & 3 \\
Repeated passenger cancellations & 1 & 1 & 1 \\
\bottomrule
\end{tabular}

\begin{tablenotes}
\item \tiny \textbf{Lik.} = Likelihood (1–5); \textbf{Imp.} = Impact (1–5); \textbf{Score} = L$\times$I.
\end{tablenotes}
\end{threeparttable}
\end{table}

Through the risk scoring framework, we ranked the anomalies based on their severity, likelihood, and impact on the real world as shown in the previous table (Table~\ref{tab:table2}).

\begin{figure}[!t]
    \centering
    \includegraphics[width=\linewidth]{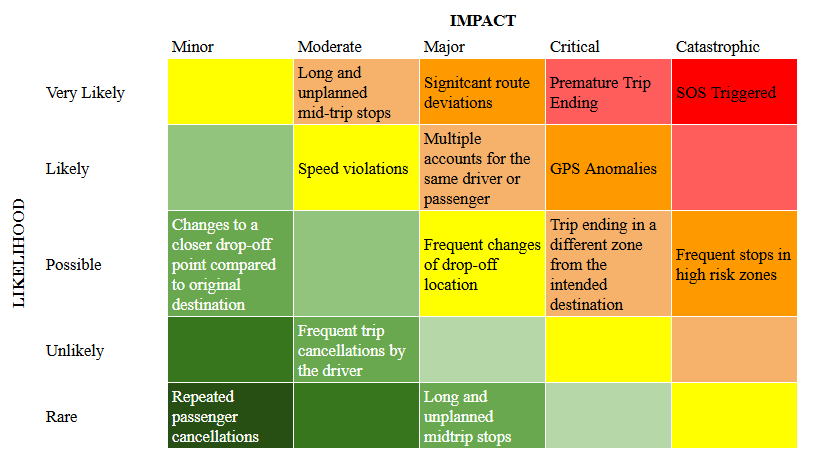}
    \caption{Risk Matrix}
    \label{fig:fig2}
\end{figure}

This ranking informed the weighted risk matrix that combines impact scores (1–5) with likelihood estimates, validated using historical incident data from Nairobi and Dar es Salaam shown in Fig.~\ref{fig:fig2}.

\subsubsection{System Implementation}
For the scope of this research, we focus exclusively on premature trip endings as the anomaly of interest. When a trip ends prematurely, our system initiates a post-track sequence called Proximity Sync, which monitors the relative distance between the driver and passenger devices to provide forensic traceability beyond the official trip termination.

The system is implemented as middleware, a lightweight Flask-based web server that acts as the core processing unit. It exposes REST API endpoints that accept JSON-formatted trip logs from client applications like our simulated Flutter app. Upon receipt, the middleware parses the data, stores it in a relational database, and applies the post-track logic. Processed results are made accessible through both API endpoints and a simple web-based dashboard. We used the following technologies:
\begin{itemize}
    \item Python (Flask) for the API server and dashboard rendering.
\item SQLite with SQLAlchemy ORM for database persistence.
\item Chart.js for distance-over-time visualization.
\item Leaflet.js (OpenStreetMap) for map visualization.
\item HTML/Jinja2 templates for the minimal web-based dashboard.
\end{itemize}

\subsubsection{Post-Track Engine}
Post-Track Engine is the forensic center of the system. During an abnormality detection (e.g., trip cancellation before destination), there is temporary synchronization between passenger device and driver device. Exchanging location data is performed every 30 seconds through an encrypted channel. The distance of the two users is calculated by using the Haversine formula, refer to the Equation (\ref{eq:haversine}) below, and if the users remain within 10 meters of distance from each other for some duration, the system logs the event as a post-trip event.

\begin{equation}
D = 2r \arcsin \sqrt{ \sin^2 \left( \frac{\Delta \phi}{2} \right) + \cos(\phi_1)\cos(\phi_2)\sin^2 \left( \frac{\Delta \lambda}{2} \right) }
\label{eq:haversine}
\end{equation}

where $r$ is the Earth's radius, $\phi_1$ and $\phi_2$ are the latitudes, and $\lambda_1$, $\lambda_2$ are the longitudes of the two users (driver and passenger), all in radians. 

\textbf{Threshold Justification:}
25 meters were chosen as they reflect common GPS accuracy ranges, which can vary between 5 and 20 meters, and can possibly exceed in dense urban environments. A 10 meter threshold is used to simulate when a passenger is still inside, or physically close to the vehicle, while a 25-meter distance approximates separation scenarios. These thresholds are scalable and tuned in subsequent deployments for empirical calibration and conditions of the environment. This forensic log provides investigators with evidence of continued proximity, improving trace reconstruction. Data are anonymized and encrypted before storage.

\subsubsection{Post-Track Logic}
\begin{figure}[!t]
  \centering
  \includegraphics[width=0.9\columnwidth]{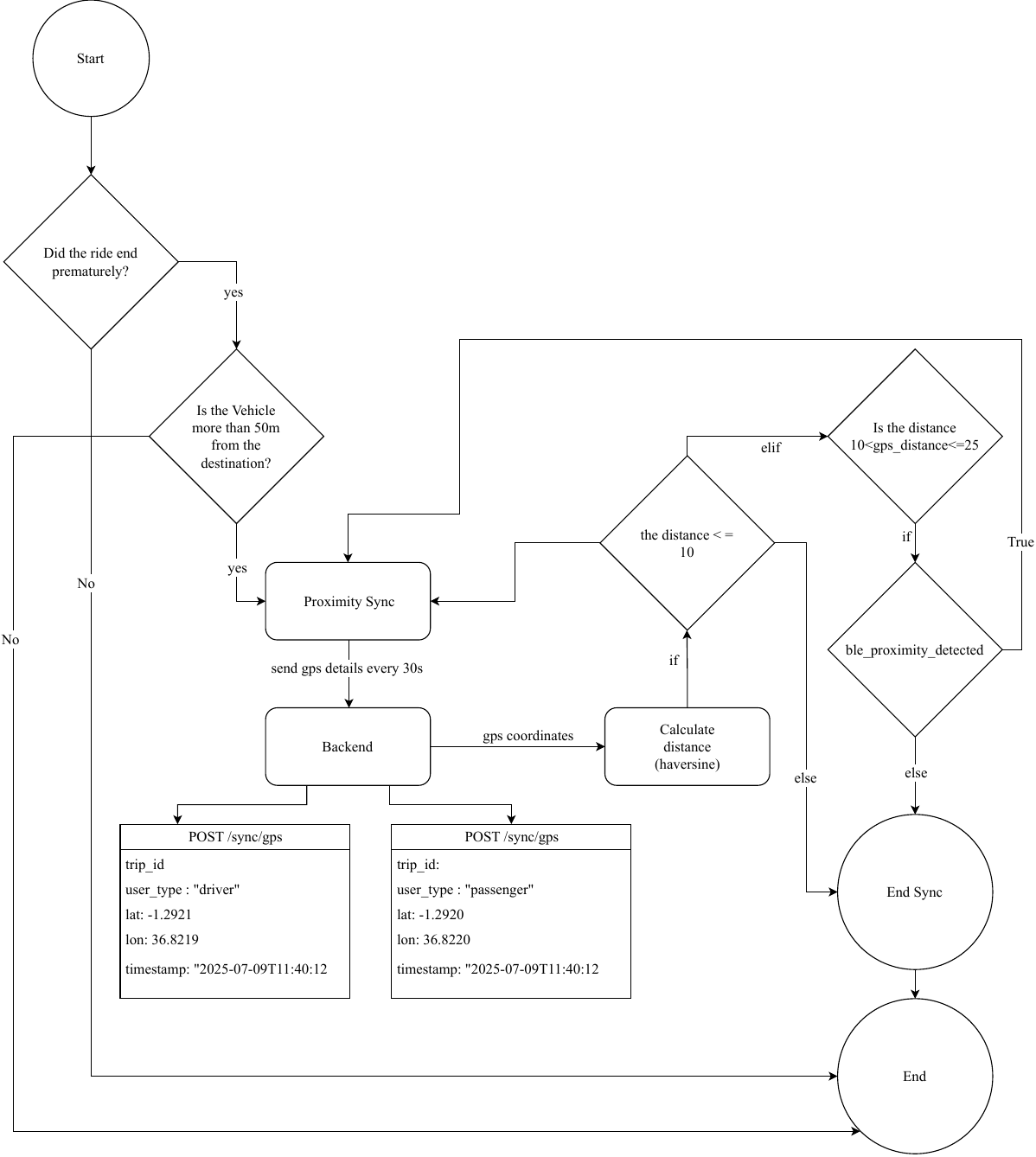}
  \caption{Process Flowchart for Post Track Sync}
  \label{fig:post_track_flow}
\end{figure}

To enhance forensic traceability in the event of premature trip termination, as illustrated in Fig~\ref{fig:post_track_flow}, the system initiates a post-track monitoring sequence called Proximity Sync, whose algorithmic flow is shown in Listing 1.

Listing 1. Post-Track Synchronization Algorithm

\begin{lstlisting}[language=Python, 
                   basicstyle=\ttfamily\scriptsize, 
                   backgroundcolor=\color{gray!10}, 
                   frame=single, 
                   numbers=none, 
                   xleftmargin=3pt, 
                   breaklines=true, 
                   breakatwhitespace=true, 
                   postbreak=\mbox{\textcolor{gray}{$\hookrightarrow$}\space}, 
                   aboveskip=2pt, 
                   belowskip=2pt]

procedure POSTTRACK(TRIP_ID):
    if isPrematureEnd(TRIP_ID) and FartherThan(TRIP_ID, 50):
        startTimer(300)  # 5 min retention
        while timerActive():
            gp_d = getGPS("driver")
            gp_p = getGPS("passenger")
            d = haversine(gp_d.lat, gp_d.lon, gp_p.lat, gp_p.lon)

            if d <= 10:
                logEvent(TRIP_ID, "end_sync_close", d, now())
                break
            elif 10 < d <= 25 and bleProximityDetected(TRIP_ID):
                logEvent(TRIP_ID, "end_sync_ble", d, now())
                break

            logSample(TRIP_ID, d, gp_d, gp_p, now())
            sleep(30)

        if timerExpired():
            logEvent(TRIP_ID, "sync_timeout_or_stopped", d, now())
end procedure
\end{lstlisting}

This feature is designed to determine whether the passenger remains in physical proximity to the vehicle, and thus the driver, after a trip has been forcefully ended or canceled. When a trip ends prematurely and the endpoint is more than 50 meters from the original destination, the system creates a new record containing anonymized driver and passenger identifiers and timestamps the start of the sync. Both devices periodically transmit GPS coordinates to the \texttt{/location} endpoint, and for each new driver–passenger location pair collected within one minute of each other, the system computes the real-time distance using the Haversine formula. These distances are logged in the database with corresponding timestamps. Once the computed distance exceeds a configurable threshold of 50 meters by default, the system concludes the Proximity Sync, records the separation time, and updates the trip status to “separated.” This implementation ensures that after an early termination, Safe2Hail maintains a continuous digital trail indicating whether the driver and passenger remained physically proximate. This post-track flow increases the forensic resolution of trips and allows anomaly review boards or platform security teams to reconstruct what happened after trip cancellation.

\subsubsection{Dashboard} While maintaining immutable and secure logs, ride-hailing service providers need access to live operational data and real-time post track trace. This capability enables monitoring of trip patterns, early safety concern detection, and preemptive action without compromising driver or passenger privacy. To accomplish this, we developed a web-based dashboard interfacing with the middleware layer. This dashboard will provide a visual and interactive interface that aggregates and projects key metrics related to trip behavior and post-ride trace. The main components will include the trip information. 
This section is a tabular view of all ongoing trips, including the Trip ID, Driver's ID, Passenger's ID, and status. The dashboard is designed with modularity in mind, allowing additional data layers to be integrated in future iterations. Potential future enhancements include real-time alerts, anomaly trend analytics, and role-based access for different administrative tiers. This modular approach ensures the dashboard can evolve and expand its capabilities over time, while the post-track integration ensures continuous safety oversight even after standard ride-hailing app protections are no longer active.

\section{Safe2Hail System Prototype}

\begin{figure}[!t]
\centering
\begin{tikzpicture}
\begin{axis}[
  width=0.97\columnwidth,
  height=0.43\columnwidth,
  xlabel={Time (s)},
  ylabel={Distance (m)},
  xmin=0, xmax=300,
  ymin=0, ymax=60,
  grid=both,
  grid style={dashed, line width=0.3pt},
  tick align=outside,
  tick label style={font=\scriptsize},   
  legend style={draw=none, font=\scriptsize, at={(0.97,0.97)}, anchor=north east},
  every axis plot/.append style={line width=0.9pt}
]

\addplot+[mark=none, domain=0:300, samples=200] {50*exp(-x/80)};
\addlegendentry{Proximity distance}

\addplot+[mark=none, dashed] coordinates {(0,25) (300,25)};
\addlegendentry{25 m threshold}

\addplot+[mark=none, dashed] coordinates {(0,10) (300,10)};
\addlegendentry{10 m threshold}

\addplot+[only marks, mark size=1pt]
table[row sep=\\]{
x   y
0   50.0\\
30  39.7\\
60  31.5\\
90  25.0\\
120 19.8\\
150 15.7\\
180 12.4\\
210 9.8\\
240 7.8\\
270 6.2\\
300 5.0\\
};
\addlegendentry{30 s samples}

\end{axis}
\end{tikzpicture}
\caption{Distance vs. time showing post-trip proximity persistence and decision thresholds (10 m and 25 m).}
\label{fig:graph}
\end{figure}
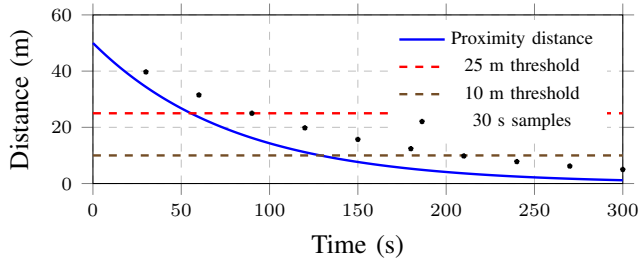

We developed a working prototype of the Safe2Hail middleware and dashboard to validate the feasibility of post-trip proximity monitoring. The system operates through several coordinated components, including a driver app simulator and a passenger app simulator. 
Upon trip termination, a POST request is sent to the \texttt{/trip/end} endpoint, which initializes a synchronization record and sets the trip status to syncing. During real-time proximity monitoring, the \texttt{/location/<trip\_id>} endpoint receives periodic GPS updates from both driver and passenger devices. The system stores these location logs and, when both are recent, computes their spatial distance. The calculated distances are recorded in the DistanceLog table and simultaneously visualized through the dashboard interface, as illustrated in Fig.~\ref{fig:graph}, which shows the temporal evolution of proximity distance alongside decision thresholds.
All trip metadata, location logs, and computed distances are stored in a SQLite database. A structured JSON log accessible through a GET request to \texttt{/trip/<trip\_id>/log}, which contains key details such as synchronization start and end times, duration, final status, and the complete time-series distance record. 

The dashboard comprises two main views: the Index View, which lists all trips with their IDs, driver and passenger identifiers, and current status whether syncing or separated; and the Trip Detail View, which presents a line chart, implemented with Chart.js.

For testing and simulation, a minimal Flutter application was developed to emulate trip initiation, premature termination, and periodic GPS updates. This application generated test logs, Table~\ref{tab:sim}, that were used to validate the middleware’s ability to accurately transition trip statuses from syncing to separated.

\textbf{Evaluation Metrics Clarification:} The accuracy values reported in Table III represent the system’s ability to correctly classify post-trip proximity outcomes, specifically distinguishing between “proximity maintained” and “separation” events. Ground truth labels were derived from controlled simulation scenarios in which the spatial relationship between driver and passenger was predefined. The system’s classification outputs were compared against these predefined ground truth outcomes. While this provides an initial validation of the framework’s logical correctness, future work will incorporate real-world deployments and statistical validation using empirical datasets.

\begin{table}[!b]
\centering
\caption{Simulation Results Summary}
\scriptsize 
\setlength{\tabcolsep}{3.5pt} 
\renewcommand{\arraystretch}{1.05} 
\begin{tabular}{lccc}
\toprule
\textbf{Scenario} & \textbf{Duration (s)} & \textbf{Accuracy (\%)} & \textbf{Status}\\
\midrule
Premature End & 45 & 95 & Proximity logged\\
Cancellation & 38 & 90 & Sync terminated\\
Route Deviation & 52 & 92 & Alert issued\\
GPS Disabled & 48 & 89 & Trace lost\\
\bottomrule
\end{tabular}
\label{tab:sim}
\end{table}

\section {Discussion}
The Safe2Hail prototype demonstrates the feasibility of post-trip proximity monitoring as a complementary forensic mechanism within ride-hailing ecosystems. Unlike existing anomaly detection methods that cease operation once a trip is canceled or prematurely terminated, Safe2Hail continues to monitor and log driver–passenger proximity until physical separation is confirmed. This continuous post-trip tracking offers several notable advantages.

First, it enhances forensic traceability, allowing investigators to determine whether a passenger remained in the vehicle after an abrupt ride termination. Second, it upholds a privacy-aware design, as the system stores only anonymized identifiers and distance data without capturing personal information. Third, its lightweight and adaptable architecture, implemented using Flask and SQLite, ensures that the middleware can be seamlessly integrated into existing ride-hailing infrastructures without significant overhead.

Despite these benefits, the current prototype has several limitations. The system relies solely on GPS-based logging, which can be imprecise in dense urban environments where signal accuracy is reduced. Additionally, the absence of Bluetooth Low Energy (BLE) or Near Field Communication (NFC) proximity verification constrains the system’s ability to confirm “same-vehicle” presence with higher accuracy. The dashboard design is also intentionally minimalistic, reflecting its proof-of-concept nature rather than a production-ready user interface.

Overall, Safe2Hail addresses an unmitigated gap in ride-hailing security—passenger safety following premature trip termination. By extending monitoring beyond the official end of a ride, it introduces an additional layer of accountability and contributes to building greater trust and transparency within 
ride-hailing platforms.
Table~\ref{tab:compare} highlights how Safe2Hail extends forensic capability relative to existing safety features.

\begin{table}[h]
\centering
\caption{Comparison with Existing Ride-Hailing Safety Features}
\begin{tabular}{lcc}
\toprule
\textbf{Feature} & \textbf{Traditional Apps} & \textbf{Safe2Hail}\\
\midrule
In-trip Tracking & Active only & Extended post-trip\\
SOS Alert & Manual & Automatic anomaly-based\\
Data Retention & Limited & 24-hour forensic log\\
Privacy Model & Centralized & Encrypted, temporary\\
\bottomrule
\end{tabular}
\label{tab:compare}
\end{table}

This comparison underscores that while existing systems terminate monitoring at trip completion, Safe2Hail uniquely continues forensic logging in a privacy-aware and time-bounded fashion.

\begin{samepage}

\textbf{Limitations:}
While it is effective, the existing mechanism relies on the assumption that all devices involved behave cooperatively, meaning both driver and passengers’ devices do not turn off and actively transmit their location information. However, in adversarial circumstances such as when GPS is purposefully disabled or a device is compromised, this proximity detection can be circumvented. Furthermore, the current evaluation is based on modeling and does not fully account for the range of variability shown by real-world deployment conditions, such as network instability or GPS noise in urban canyons. It is envisaged that these limitations will be addressed through real-world deployments and the integration of more robust proximity sensing mechanisms.

\end{samepage}

\section {Future Work}
Future development of the Safe2Hail system will focus on enhancing proximity detection accuracy and expanding its real-world applicability. Integration of Bluetooth Low Energy (BLE) or Ultra-Wideband (UWB) sensors will allow the system to confirm short-range proximity more precisely, mitigating GPS inaccuracies in dense urban areas. Additionally, a cloud-based data pipeline and asynchronous messaging system could improve scalability for large-scale ride-hailing deployments. Further research will also explore integrating Safe2Hail with emergency response systems for automated alerts when abnormal proximity patterns persist. Finally, applying blockchain or distributed ledger technologies can provide immutable post-trip logs, ensuring tamper-proof forensic evidence for regulatory and investigative purposes.

\section {Conclusion}
This research presents Safe2Hail, a novel forensic-driven framework for trip-level anomaly detection in ride-hailing platforms, specifically addressing safety challenges in urban contexts such as Dar es Salaam and Nairobi. By prioritizing critical trip anomalies through systematic risk assessment and implementing a post-trip Proximity Sync mechanism for premature trip endings, the system bridges the gap between in-app safety features and post-incident investigation needs. The developed middleware and dashboard prototype demonstrate that proximity-based monitoring can provide forensic traceability and enhance accountability while preserving user privacy. Built using a lightweight Flask-based architecture and GPS-driven analytics, Safe2Hail extends safety oversight beyond the official trip duration, offering a scalable and adaptable solution for diverse ride-hailing ecosystems. Ultimately, this work contributes to stronger post-incident digital evidence, improved transparency, and greater passenger and driver safety across different operational and geographical contexts.

\bibliographystyle{IEEEtran} 
\bibliography{references}
\balance

\end{document}